\documentclass[prd,twocolumn,showpacs,amsmath,amssymb,nofootinbib]{revtex4}

\newcommand\lsim{\lesssim}
\newcommand\gsim{\gtrsim}

\newcommand\vev[1]{{\langle {#1} \rangle}}
\renewcommand\({\left(}
\renewcommand\){\right)}
\renewcommand\[{\left[}
\renewcommand\]{\right]}

\newcommand\eq[1]{Eq.~(\ref{#1})}
\newcommand\eqs[2]{Eqs.~(\ref{#1}) and (\ref{#2})}
\newcommand\eqss[3]{Eqs.~(\ref{#1}), (\ref{#2}), and (\ref{#3})}

\newcommand\eqst[2]{Eqs.~(\ref{#1})--(\ref{#2})}

\newcommand\eqreff[1]{(\ref{#1})}

\newcommand\pa{\partial}

\newcommand\ee{\end{equation}}
\newcommand\be{\begin{equation}}
\newcommand\eea{\end{eqnarray}}
\newcommand\bea{\begin{eqnarray}}

\newcommand\mpl{M_{\rm P}}

\def\calp{{\cal P}}

\newcommand\bfk{{\mathbf k}}

\newcommand\bfp{{\mathbf p}}
\newcommand\bfq{{\mathbf q}}

\newcommand\bfx{{\mathbf x}}
\newcommand\bfy{{\mathbf y}}

\newcommand\sub[1]{_{\rm #1}}
\newcommand\su[1]{^{\rm #1}}

\newcommand\mone{^{-1}}

\newcommand\mthree{^{-3}}

\newcommand{\fnl}{f\sub{NL}}

\newcommand{\tnl}{\tau_\zeta}

\newcommand\bfkp{{{\bfk}'}}

\newcommand\tpq{{(2\pi)^3}}

\newcommand\ksmooth{k\sub{max}}
\newcommand\lmone{L^{-1}}
\newcommand\mmone{M^{-1}}

\newcommand\calps{\calp_\sigma}
\newcommand\calpz{\calp_\zeta}

\newcommand\pps{P_\sigma}
\newcommand\ppz{P_\zeta}
\newcommand\ppzm{P_{M\zeta}}
\newcommand\tilpps{\tilde P_\sigma}
\newcommand\tilppz{\tilde P_\zeta}
\newcommand\tilppzm{\tilde  P_{M\zeta}}

\newcommand\dsig{\delta\sigma}
\newcommand\dsigs{\delta\sigma^2}
\newcommand\sigbar{ \overline\sigma }
\newcommand\dsigm{\delta\sigma_M}

\newcommand\sigbarm{ \overline\sigma_M }

\begin{document}

\title{The curvature perturbation in a box} 

\author{David H.~Lyth}

\affiliation{{\it Physics Department, Lancaster University, Lancaster
LA1 4YB, UK}}

\begin{abstract}
The stochastic properties of cosmological perturbations
 are best defined through the Fourier expansion
in a finite box. I discuss the reasons for that with reference to the
curvature perturbation, and explore some issues arising from it.
\end{abstract}
\maketitle

\section{Introduction} 

Strong observational constraints on the primordial curvature perturbation 
$\zeta$ make
it an important discriminator between models of the very early Universe.
 After smoothing relevant quantities on the shortest scale of interest,
  $\zeta$  may be defined as the fractional perturbation $\delta a/a$
of the locally-defined scale factor $a(\bfx,t)$. Equivalently, it is
 the perturbation in the number $N(\bfx,t)$ of $e$-folds of
expansion, starting on an initial   `flat' slice of spacetime
(where $\delta a=0$) and ending on a
slice of uniform energy density at time $t$. The  spacetime threads
of constant $\bfx$ are taken to be comoving.

Any  choice of the initial `flat'
slice will do, as long as the smoothing scale is  outside the horizon
at that stage, because the expansion going from one `flat' slice to another
is uniform.  The final slice is to be located before the smoothing scale
 re-enters the horizon, and on cosmological scales it should be
late enough that $\zeta$ has settled down to the 
 final time-independent value that is constrained by observation.
With the smoothing scale outside the horizon, the evolution at each
point is expected to be  that of some unperturbed universe (the separate
universe assumption \cite{starob85,ss,lms,lr,mycurv}.

During inflation, one or more of the scalar field perturbations
 is supposed to be practically massless during inflation ($m^2\ll H^2$).
For these `light' fields, the vacuum fluctuation is promoted to a classical
perturbation as each scale leaves the horizon. To calculate $\delta N$
one smooths the light field perturbations on a comoving scale shorter than
any of interest. A few Hubble times after the scale leaves the horizon
the light fields have classical perturbations. Their values at this 
epoch are supposed
to determine $N(\bfx,t)$ and hence $\zeta$.  
Taking for simplicity just one field $\sigma$ we have  
\bea \zeta(\bfx,t) &=&  \delta N( \sigma(\bfx) ) \\
&\equiv&  N(\sigma(\bfx)) - N(\sigbar) \\
&=& 
 N'\dsig(\bfx)  + \frac12 N'' \dsig^2(\bfx) + \cdots 
, \label{withlin} \eea
where 
\be
\dsig(\bfx)  \equiv \sigma(\bfx) - \sigbar
, \label{vevs} \ee
and
\be N'\equiv \left. \frac{d N}{d \sigma} \right|_{\sigbar}, \qquad
N''\equiv \left. \frac{d^2 N}{d \sigma^2} \right|_{\sigbar}
\label{withlindefs} \ee
and so on. 

The final expression for $\zeta$ is a power series in the field perturbation
defined on the initial slice. 
The unperturbed value of $\sigma$  is defined as its  spatial 
average of each field, which simplifies the analysis. 

We are interested in the Fourier components of $\zeta$:
\be
\zeta_\bfk = \int e^{-i\bfk\cdot \bfx}  \zeta(\bfx) d^3x
, \ee
and similarly for $\dsig$. The  perturbation 
$\delta\sigma_\bfk$ is supposed to originate from the vacuum fluctuation,
and can be considered \cite{slyth} as classical starting a few Hubble times
after the epoch of horizon exit $k=aH\equiv \dot a$, where $a$ is the scale
factor. At this stage the correlators of $\dsig_\bfk$ on scales not too
far outside the horizon can easily 
be calculated using perturbative  quantum field
theory. Assuming Lorentz invariance and a quadratic kinetic term,
$\dsig$  is almost gaussian. The  subsequent separate-universe
 evolution of $\dsig(\bfx)$ then gives  the correlators of
$\dsig$ at the initial epoch. 
 Finally, \eq{withlin} gives the correlators
of $\zeta$ which can be compared with observation.

To avoid assumptions about the unknownable Universe very far beyond the 
present horizon $H_0\mone$, this calculation should be done within a 
comoving box,  whose present size $L$ is not too much bigger than 
$H_0\mone$.  This situation has not been discussed much in the literature 
for two reasons. First, with the box size not too much bigger
 than $H_0\mone$ (minimal box)  the linear term of \eq{withlin}
dominates. Then  correlators   
 are practically independent of
the box size, so that only  $\sigbar$ need be specified.
Second, if $\sigma$ is the inflaton in single-field slow-roll inflation
and we use a minimal box,
then $\sigbar$ can be calculated from the inflation model.

On the other hand, $\sigma$ may have nothing to do with inflation, as in
the curvaton model \cite{curvaton}. 
Also, we should know how to handle the dependence on box size 
as a matter of principle. In this note I look
at some of the issues raised when one takes seriously the box size, 
developing some earlier work  \cite{bl,mycurv}.

\section{Correlators of $\zeta$}

A  model of the early Universe will predict, not $\zeta(\bfx)$ itself
but   correlators, 
$\vev{\zeta(\bfx) \zeta(\bfy)}$ and so on. The $\vev{}$
is  an ensemble average, which in the inflationary 
cosmology becomes a Heisenberg-picture vacuum expectation value. 
As  the vacuum is invariant under translations, so are  the 
correlators.

The vacuum is also invariant  under rotations, and so are the 
correlators.\footnote
{In considering translations and rotations one takes $\bfx$
to denote Cartesian coordinates defined in the unperturbed universe.
In the perturbed universe these same coordinates cover the curved spacetime
slice  of fixed $t$, in a way that is defined by the threading which is
here taken to be comoving.}
Invariance under translations and rotations constrains the form
 of the  correlators. One usually considers only the two,
three and four-point correlators.

 The two-point correlator is
\be
\vev{\zeta_\bfk  \zeta_\bfkp}= \tpq \delta^{(3)}(\bfk+\bfkp ) P_\zeta(k)
\label{calpg} , \ee
where $\ppz$ is the spectrum. 
It is useful to define  
 $\calp_\zeta \equiv (k^3/2\pi^2) P_\zeta$,  also called the spectrum.
The mean-square of $\zeta$ is
\be
\vev{\zeta^2(\bfx)} = \int^{\ksmooth}_{\lmone} \calpz(k)  \frac{dk}k
, \ee
where $\ksmooth$ is the scale leaving the horizon at the initial epoch,
and the infrared cutoff is provided by the box size.
On cosmological scales observation gives $\calpz=(5\times 10^{-5})^2$ with
little scale dependence.

The three-point correlator is 
\be \vev{ \zeta_{\bfk_1} \zeta_{\bfk_2}
\zeta_{\bfk_3} } = \tpq \delta^{(3)}(\bfk_1+\bfk_2+\bfk_3)
B_\zeta(k_1,k_2,k_3) 
\label{bg}
,
\ee
where $B_\zeta$ is the bispectrum.
The  connected contribution to the four-point correlator is
\be
\vev{\zeta_{\bfk_1}\zeta_{\bfk_2}\zeta_{\bfk_3}\zeta_{\bfk_4}}\sub c=
(2\pi)^3\delta^{(3)}(\bfk_1+\bfk_2+\bfk_3+\bfk_4) T_\zeta
\label{tg}. \ee
The trispectrum   $T_\zeta$ is    a function of six scalars,  defining  the
quadrilateral formed by $\{\bfk_1,\bfk_2,\bfk_3,\bfk_4\}$.
There is also a disconnected contribution:
\be
\vev{\zeta_{\bfk_1} \zeta_{\bfk_2}\zeta_{\bfk_3}\zeta_{\bfk_4}}\sub d=
\vev{\zeta_{\bfk_1} \zeta_{\bfk_2}}\vev{\zeta_{\bfk_3}\zeta_{\bfk_4}} 
+  {\rm perms}
. \ee

For any correlator, the connected correlator is the one that comes with an
overall delta function.
If the two-point correlator is the only connected one, $\zeta$ is said to be
Gaussian.
Data are at present consistent with the hypothesis that $\zeta$
is perfectly gaussian, but they  might not be  in the future.

As a result of translation invariance, the ensemble averages
$\vev{\zeta(\bfx)\zeta(\bfy)}$ etc.\ can be regarded as a spatial
average with fixed $\bfx-\bfy$. This is the ergodic theorem,  whose proof 
for correlators is very simple. Considering the two-point correlator
the definition of $P_\zeta$ gives
\be
\vev{\zeta(\bfy)\zeta(\bfx+\bfy)}= \int P(k) e^{i\bfk\cdot\bfx} d^3k
. \label{ensemav} \ee
On the other hand, the  spatial average is
\bea
{}&& L\mthree \int \zeta(\bfy)\zeta(\bfx+\bfy) d^3y \nonumber \\
&=& 
L\mthree(2\pi)^{-6} \int \zeta_\bfk\zeta_\bfkp 
e^{i[\bfk\cdot\bfy+\bfkp\cdot(\bfx+\bfy)]} d^3y d^3k d^3k' \nonumber \\
&=& L\mthree 
(2\pi)^{-3} \int \delta^3(\bfk+\bfkp) \zeta_\bfk\zeta_\bfkp e^{i\bfk\cdot
\bfx} d^3k d^3k' 
. \label{spaceav} \eea
In the final expression, $\zeta_\bfk\zeta_\bfkp$ can be replaced by its
ensemble average  $\vev{\zeta_\bfk\zeta_\bfkp}$, 
because each volume element $d^3kd^3k'$ can contain an
arbitrary large number of of points. (Remember that we are working in a finite
box so that the possible momenta form a cubic lattice in $\bfk$-space.
Within a cell, the Fourier coefficients are uncorrelated because of the 
delta functions in \eqss{calpg}{bg}{tg} and so on.)
Writing $\vev{\zeta_\bfk\zeta_\bfkp}$ in terms of $P(k)$ and 
using the rule $[\delta^3(\bfk-\bfkp)]^2 = L^3\delta^3(\bfk-\bfkp)$, we see
that the spatial average  \eqreff{spaceav} is indeed equal to the ensemble
average \eqreff{ensemav}.
reproduces the ensemble average, and the same argument works for higher
correlators too. 

{}From this proof it is clear that the 
ergodic theorem works in a finite box, with the usual proviso that the
box size is bigger than scales of interest. Its proof relies just on 
 translation invariance, which makes mathematical 
sense in the finite box because of
the periodic boundary condition.

\section{The correlators of $\dsig$}

The correlators of $\dsig$ have the same form as those of $\zeta$.
 In particular
\bea
\vev{\sigma_\bfk  \sigma_\bfkp}&=& \tpq \delta^{(3)}(\bfk+\bfkp ) P_\sigma(k)
\label{sigspec}
\\
\vev{\sigma_{\bfk_1} \sigma_{\bfk_2}\sigma_{\bfk_3}\sigma_{\bfk_4}}
\sub d &=&
\vev{\sigma_{\bfk_1} \sigma_{\bfk_2}}\vev{\sigma_{\bfk_3}\sigma_{\bfk_4}} 
+  {\rm perms}
. \label{sigdis} \eea

The  chosen
box should be well inside the horizon at the beginning of inflation.
After a few Hubble times, the 
 universe inside the box is then expected \cite{wald} to become practically
homogeneous and isotropic at the classical level. 
The conversion of the vacuum fluctuation into a classical perturbation
$\dsig$  begins  when the box leaves the horizon at the epoch $k=aH$.

We observe scales $k\gsim H_0$ where $H_0$ is the present value of the 
Hubble parameter. Taking inflation to be almost exponential there 
 are at most  60 or so Hubble times between horizon exit
for the  scale $k=H_0$ and the end of inflation. Of most interest is the 
perturbation on cosmological scales, which  leave
the horizon during the first 10 or so Hubble times. Smaller scale perturbations
could also be of interest, for example to form black holes at the end of
inflation.

We  can distinguish between two kinds of box. A {\em minimal box}, for which
$\ln (LH_0)$ is not too big,   will leave the horizon just a few
$e$-folds before the observable Universe leaves the horizon, 
and hence not too many Hubble times before all observable scales leave the 
horizon. A   {\em super-large box} on the other hand,
 with very large $\ln (LH_0)$,  would leave the horizon very 
many $e$-folds before the observable universe. 
We will see how to calculate things in a minimal box, and poin to the 
difficulties that are encountered if one considers instead a super-large box.

With the minimal box, $\dsig$ remains small at least while cosmological
scales leave the horizon. 
On the usual assumption that $\sigma$ is canonically normalized,
the perturbation $\dsig$ generated from the vacuum is then
almost gaussian on scales not too far outside the horizon.
Its spectrum is \cite{bd}
\be
\calps(k) \simeq (H_k/2\pi)^2
, \label{flatspec} \ee
 where $H_k$ is the Hubble parameter at horizon exit. 
The bispectrum \cite{sl}
and trispectrum \cite{sls} of $\dsig$ have been calculated, and are
suppressed by slow-roll factors. (These are $|\dot H/H^2|\ll 1$, needed for
inflation, and parameters involving derivatives of the 
 potential $V(\sigma)$ which have to be small to justify the
perturbative quantum field theory calculation.)

As we saw earlier, the 
 correlation functions defined by the spectrum, bispectrum and so on
may be defined as spatial averages within the box. With a minimal
box it is reasonable to assume that these are the same as the  spatial
averages within the region of size $H_0\mone$ around us, that can 
actually be observed.


If we consider instead a super-large box several difficulties arise.
We have to assume that enough inflation took place for the box to exist,
and we need to understand the field theory starting from the era when the
box leaves the horizon. When  observable scales leave the horizon,
the perturbation that has already been generated on larger scales my be
large, which would complicate the calculation of $\delta\sigma$.
And even when the calculation has been performed, the spatial averages
represented by the correlators may have nothing to do with the spatial
averages that are actually observed.

\section{The mean value $\sigbar$}

\subsection{With $\sigma$ the inflaton}

Now we come to the mean value $\sigbar$, of the field within the box.
If $\sigma$ is the inflaton in a single-field slow-roll inflation model,
and we use a minimal box, then $\sigbar$ when $N$ $e$-folds of inflation 
remain is given by
\be
N(\sigma) = \mpl^2 \int^{\sigbar}_{\sigma\sub{end}} \frac V {V'} d\sigma
. \label{neq}\ee
This follows  from $dN=Hdt$ and  the  slow-roll approximations
\bea
3H\dot\sigma &=& -V' \\
3H^2\mpl^2 &=& V \\
\dot H/H^2 &=& -\dot\sigma^2
.\eea
The inflation model will give the field $\sigma\sub{end}$ at the end of 
inflation, and the post-inflationary cosmology determines $N$ when a given
scale leaves the horizon, with $N\sim 50$ or so for cosmological scales in 
the usual cosmology. 

In this very special case, $\zeta$ becomes time-independent soon after
horizon exit. As a result the
 small perturbation $\zeta=\delta N\sim 10^{-5}$  is then
given  in terms of the potential and its derivatives \cite{sl,bsw}:
\bea
\zeta &=& \frac1{\mpl^2}\frac{V}{V'} \delta \sigma 
+ \frac12 (2\eta-\epsilon) \(\frac1{\mpl^2}\frac{V}{V'} \delta \sigma  \)^2
 \nonumber \\
&+& \frac16 (2\epsilon \eta - 2\eta^2 + \xi^2) 
\(\frac1{\mpl^2}\frac{V}{V'} \delta \sigma  \)^3 + \cdots
, \label{slowroll} \eea
where 
$2\epsilon=\mpl^2(V'/V)^2$, $\eta=\mpl^2V''/V$ and 
$\xi^2= \mpl^4 V'''V'/V^2$.
With a minimal box the linear term dominates, and
 taking the 
initial slice to be soon after horizon exit on a given scale we then arrive
at the famous result
\be
\calpz \simeq \( \frac1{\mpl^2} \frac V{V'} \frac H{2\pi} \)^2
,\ee
where the right hand side is evaluated at horizon exit. The bispectrum and
trispectrum can also be calculated, as described after \eq{ratios}. 

Going to a super-large box, none of this may work. As we noted earlier,
$\delta\sigma$ might be big which would invalidate the calculation of its
stochastic properties. Also, $\delta N$ might be big and then \eq{neq}
would apply only to the average of $N$ within the box ($\sigma\sub{end}$
then being a spatial average),
which might have little to do with the situation in the observable Universe.

\subsection{With $\sigma$ a curvaton-type field}

Now suppose instead that $\sigma$  has nothing to do with the inflation
dynamics, as is typically (though not inevitably \cite{gong})
the case in the curvaton model.\footnote
{For slow-roll inflation  this corresponds to $V'$
being much less than the corresponding quantity for the inflaton but 
we are not making  any assumption about the inflation model.}
Then at last we encounter a case where it may be useful to consider a 
super-large box, using what is often called the stochastic approach to
the evolution of perturbations \cite{stochastic}.

In the stochastic approach one takes spacetime to be unperturbed with 
constant $H$ (de Sitter spacetime). One smooths 
 $\sigma$ on a practically-fixed 
scale $(1+b)H$ with $0<b\ll 1$ a constant,  and  considers the 
probability $F(t,\sigma) d\sigma$ that $\sigma(\bfx,t)$ lies within a given 
interval. It satisfies the Fokker-Planck equation 
\be
\frac{\pa F}{\pa t} = \frac{V'}{3H} \frac{\pa F }{\pa\sigma}
+\frac{H^3}{8\pi^2}\frac{\pa^2 F}{\pa \sigma^2}
, \ee
This equation, applying to any slow-rolling field, corresponds to the 
Langevin equation describing the classical evolution  plus
the random walk $\pm H/2\pi$ per Hubble time coming from the creation of
the classical perturbation from the vacuum fluctuation. 

The point now is that the probability   distribution may lose memory of the
initial condition. In particular, if
 $\dot H/H^2$ is sufficiently small, $F$ will  settle down to \cite{sy}
\be
F = \mbox{  \rm const } \exp\( -\frac{8\pi^2}{3H^4} V(\sigma) \)
\label{fprob}
 ,
\ee
More generally, one can handle a significant variation of $H$ within a
given inflation model (see for instance \cite{ouraxion,lm}).

Reverting now to a minimal box, the probability distribution $F$ should
apply to $\sigbar$ if the initial epoch is taken to be soon after the
shortest cosmological scale leaves the horizon, 
since the box size is  then not too  many $e$-folds bigger than
 the Hubble scale at the initial epoch. In any case, one could calculate
the probability distribution of $\sigbar$ by going back to the Langevin
equation. The very simplest case  arises if  $\sigma$ is a 
pseudo Nambu-Goldstone boson (PNBG) with $V'$ negligible.
Then it is defined only in some  interval  $0< \sigma < f $, and the noise
term gives $\sigma$ an equal probability of being anywhere in the interval.

Finally, given a probability distribution for $\sigbar$ one may  suppose
that the actual value for a minimal box around the observable Universe
is not too far from the most probable value (or from say the mean-square
if we are not dealing with a PNGB). Of course this final step is speculative
and may be modified by environmental considerations. Still, it seems
that in this case the use of a super-large box {\em purely to get a handle
on the likely value of $\sigbar$ within a minimal box} may be helpful.

\section{Calculating the correlators of $\zeta$}

Using the convolution theorem, 
\be
(\dsig^2)_\bfk = \int \dsig_\bfq \dsig_{\bfk - \bfq} d^3 q
, \label{conv} \ee
 \eq{withlin} determines the correlators of $\zeta$
in terms of those of $\dsig$. There is a sum of terms and the calculation
is best done using Feynman-like graphs \cite{bl,zrl,bksw}. A
 complete set of  rules for constructing the graphs is 
 given in \cite{bksw}. 
A graph with $n$ loops involves an integration of $n$ momenta, while a
tree-level involves no integration. 

Here I recall the estimates of the correlators made in \cite{bl,mycurv}
for the quadratic  truncation of \eq{withlin}. To get an idea of what
happens with higher terms included, I then consider the cubic truncation.
In both cases I take $\dsig$ to be gaussian. (See \cite{zrl} for a loop
contribution involving the bispectrum of $\dsig$.)

\subsection{Quadratic truncation}

Truncating   the field expansion after the quadratic term we have 
\be
\zeta(\bfx) =   N' \dsig(\bfx)  + \frac12 N''
\dsig^2(\bfx) 
. \label{case} \ee
There are tree-level and one-loop contributions to the correlators of 
$\zeta$:
\bea 
\ppz\su{tree} &=& N'^2 \pps(k)  \\
\ppz\su{loop}&=& 
\frac{N''^2}{(2\pi)^3}    \int_{\lmone} d^3p
\pps(p)\pps(|\bfp-\bfk|) 
\label{psigs} \\
B_{\zeta}\su{tree} &=& 2 N'^2 N''
  \pps(k_1)  \pps(k_2) + { \rm cyclic}  \label{btree} \\
B_{\zeta}\su{loop}&=&  \frac{N''^3}{(2\pi)^3}   \int_{\lmone}
d^3p  \pps(p)\pps(p_1) \pps(p_2) \label{bsigs} \\
T_{\zeta}\su{tree} &=&  N'^2 N''^2
  \pps(k_1) \pps(k_2) \pps(k_{14}) 
+  23 {\rm perms.}  \label{ttree} \\
T_{\zeta}\su{loop}&=&  \frac18  \frac{N''^4}{(2\pi)^3}  \int_{\lmone}  d^3p 
\pps(p)\pps(p_1) \pps(p_2)\pps(p_{24}) \nonumber \\
&+&  23 {\rm perms} \label{tsigs}
.
\eea
We have defined  $p_1\equiv |\bfp-\bfk_1|$, $p_2\equiv |\bfp+\bfk_2|$,
$p_{24}\equiv |\bfp + \bfk_{24}|$ and $k_{14}=|\bfk_1+\bfk_4|$. 

The 24 terms in \eq{ttree} are actually 12 pairs of identical terms,
 and the  24 terms in \eq{tsigs} are actually 3 octuplets of identical
terms.  The  tree-level contribution to the bispectrum
  was given in \cite{spergel} and the tree-level contribution to the 
trispectrum was given in \cite{bl} (see also \cite{okamoto}). 
The loop contributions to  the spectrum, bispectrum and trispectrum were
given in \cite{bl}, using $\calps$ instead of $\pps$ and with 
 $\delta\sigma$ normalized to make $N''=1$.\footnote
{The $\delta N$ formula was not invoked there, and indeed is irrelevant
in the present context. All we really need is that $\zeta$ is some quadratic
function of a gaussian quantity $\delta\phi$ with zero mean.}
 (See also  \cite{myaxion} for the loop contribution
to the spectrum of the axion isocurvature perturbation, given by an identical
formula.) 

The subscript on the integral reminds us that $\pps$ is set equal to zero
at $k<\lmone$, cutting out a sphere around each of the singularities.
This is necessary, because with $\calps$ perfectly flat there is a logarithmic
divergence whenever the argument of $\pps$ goes to zero, ie.\ in the infrared.
Allowing for slight scale dependence of $\calps$, 
 infrared convergence is slow if it occurs at all.
In contrast, there
is good convergence in the ultra-violet for any reasonable behaviour of
$\calps$, and the integral will be insensitive to the actual cutoff
 $\ksmooth$.

It is convenient to define what one might call a reduced bispectrum
$\fnl$ and a reduced trispectrum $\tnl$ by
\bea
 B_\zeta
&=& \frac65\fnl \[ P_\zeta(k_1)  P_\zeta(k_2) + {\rm cyclic} \]
\label{fnldef}  \\
T_\zeta &=& \frac12\tnl P_\zeta(k_1) P_\zeta(k_2) P_\zeta(k_{14})
 + 23 {\rm perms.}
 . \eea
At  tree-level the reduced quantities are momentum-independent:
\bea
\frac65 f\sub{NL}\su{tree} &=&  \frac{N''}{N'^2} \\
\frac12\tnl\su{tree} &=& \( \frac65 f\sub{NL}\su{tree} \)^2
. \eea

In first order perturbation theory, this
 definition of $\fnl$ coincides with the original
one \cite{spergel} in first-order cosmological  perturbation theory,
where $\fnl$ was defined with respect to the Bardeen potential
which was taken to be $\Phi= \frac35 \zeta$. In that reference it was supposed
to be independent of the momenta.
Following  \cite{maldacena} we will allow momentum dependence, and make the
definition without invoking first-order cosmological perturbation 
theory.\footnote
{We have no need of perturbation theory before horizon entry but it
is needed to evolve perturbations afterward.  Second order perturbation theory
 will  be needed if $|\fnl|\lsim 1$, and at that order
$\Phi$ and $\zeta$ are completely different functions. As a result
$\fnl$ defined with respect to $\Phi$ has nothing to do with the
$\fnl$ of the present paper.
Unfortunately,
both definitions are in use.} The quantity
here denoted  as $\tnl$ was introduced in \cite{bl} and denoted 
as $\tau\sub{NL}$.
Taking them to be momentum-independent 
observation gives bounds $|\fnl|\lsim 100$ and $\tnl\lsim 10^4$,
which on cosmological scales 
are not expected to alter much if there is momentum dependence.

Now I consider estimates of the loop contributions, taking $\calps$
to be perfectly flat and  considering only the physical regime
$k_i\gg \lmone$.
One can arrive at an estimate of $\calpz\su{loop}$
 by assuming that the integration is dominated by  spheres
around each of the two singularities,  with radii of order $k$.
This gives \cite{bl}
\be
\calpz\su{loop} \simeq  2 N''^2 \calps^2 \ln(kL)
. \ee
We deduce that
\be
\frac{\calpz\su{loop}}{\calpz\su{tree}}
  \sim  \frac{N''^2}{N'^4} \calpz  \ln(kL) 
\lsim 10^{-5} \ln(kL)
, \label{specloop}\ee
where the inequality comes from the observed spectrum
$\calpz=(5\times 10^{-5})^2$ and the observational bound on 
$\fnl$ or $\tnl$ (by coincidence those bounds give a similar result). 

The integration for $\calpz\su{loop}$ can actually be done analytically
 \cite{myaxion} and it happens to give exactly this result.
The loop integrals for the bispectrum and trispectrum cannot be done
analytically, but they can be estimated in the same way as for
$\calpz\su{loop}$. 
Although a more general estimate could be made, we will take  all
$k_i$ to be of order a common value $k$ for the bispectrum,
and all $k_i$ and $k_{ij}$ to be of order a common value for
the trispectrum. Then, estimating the loop integrals 
from the contributions of the singularities as we did for $\calpz\su{loop}$,
on finds that $(\fnl\su{loop}/\fnl\su{tree})$ and
$(\tnl\su{loop}/\tnl\su{tree})$ are both of the same order as
$(\calpz\su{loop}/\calpz\su{tree})$. Special configurations of the momenta
will give  additional factors, but observation can probe only a 
fairly limited range of momenta. Pending further investigation of that 
issue, we conclude that the loop contributions are very suppressed for a
minimal box, and unlikely to be observable.

Our general finding that the loop contributions are  negligible relies on
the observational bounds on non-gaussianity, and holds because we 
assumed that $\dsig$ is the only  field perturbation contributing to
$\zeta$. As we have seen though, in the particular case that $\sigma$
is the inflaton in a single-field slow-roll inflation model, the loop
contribution is very small   by virtue of slow-roll, and in consequence
the non-gaussianity is very small. With this in mind, we can suppose
\cite{bl}  that the inflaton field perturbation gives the dominant
contribution to $\zeta$ and write
\be
\zeta = \zeta\sub{inf} +
 N'\dsig(\bfx)  + \frac12 N'' \dsig^2(\bfx) + \cdots ,
\ee
where $\dsig$ is say the curvaton. Then the spectrum of 
$\zeta$ is dominated by that of the first term, but its bispectrum and
trispectrum  might instead be dominated by the third term which would mean that
they were dominated by a loop contribution \cite{bl}. In that case
$\zeta\sub{inf}$ could account for up to   $90\%$   of the total, without
violating observational bounds on non-gaussianity.

\subsection{Cubic truncation}

Now we include the cubic term:
\be
\zeta(\bfx) =  N' \dsig(\bfx)  + \frac12 N''\dsig^2(\bfx) +
\frac16 N'''\dsig^3(\bfx)
. \label{case2} \ee
There are no additional tree-level diagrams for the spectrum and
bispectrum, but for the trispectrum \cite{bsw} 
the cubic truncation gives  an additional term
\be
T_\zeta\su{tree3}
=  N'^3 N''' \pps(k_2)\pps(k_3)\pps(k_4)  +   \mbox{\rm 3 perms}
. \ee

The new term cannot cancel the old one since its dependence
on the momenta is quite different. To extract optimal information from
the observations one should re-define $\tnl$ and introduce a new quantity
$g_\zeta$ by writing
\bea
T_\zeta &=&
 \frac12 \tnl \ppz(k_1)\ppz(k_2)\ppz(k_{14}) + 23 \mbox{\rm \ perms.} 
\nonumber \\
&+& g_\zeta \[ \pps(k_2)\pps(k_3)\pps(k_4)  +   \mbox{\rm 3 perms}
\]
. \eea
(This $g_\zeta$ is the same as the
$g\sub{NL}$ of \cite{bsw} up to a numerical factor.)
 As with $\tnl$ it is helpful
to allow $g_\zeta$ to be momentum-dependent because it can then  correspond to
a (slightly) momentum-dependent loop contribution. 

Pending the appropriate observational analysis, we will take
 $k_i\sim k_{ij}\sim k$ and keep the original definition of $\tnl$.
Then 
\be
\tnl\su{tree3} \sim \frac{N'''}{N'^3} \ln(kL) \lsim 10^{4}
, \label{tree3} \ee
where the final inequality assumes a minimal box and uses the observational
bound on $\tnl$ which we take to be the same as if $\tnl$ were 
momentum-independent. 

Denoting the old contribution by $\tnl\su{tree2}$ we have the ratios
\be
\fnl\su{tree}:\tnl\su{tree2}:\tnl\su{tree3}
 \sim \frac{N''}{N'^2} : \( \frac{N''}{N'} \)^2 : \frac{N'''}{N'^3} 
\label{ratios}
. \ee
As pointed out in \cite{bsw}, $\tnl\su{tree3}$  might be the first
signal of non-gaussianity. This could happen in the curvaton model
if the curvaton field evolves strongly after inflation.

It could also in principle happen
if $\sigma$ is
the inflaton in a single-component slow-roll inflation model.
Slow-roll requires only that all three slow-roll parameters 
are $\ll 1$, and  one might have over a limited range of scales
$|\xi|^2\gg |\eta|$ and $|\xi|^2\gg \epsilon$.
In such a  case though, we have to remember that the small non-gaussianity
of $\dsig$ at horizon exit will be comparable with the non-gaussianity
that we are considering here (ie.\ that generated by the non-linearity
of the $\delta N$ formula).  
The known estimates of  the bispectrum 
\cite{sl} and
trispectrum \cite{sls} of $\dsig$ at horizon exit
assume that $\xi^2$ is negligible, and will require modification if it
is not. Note also that in such a case
the usual \cite{llcobe}  formula
 $n-1=2\eta-6\epsilon$ the spectral index may fail as well
\cite{ewangsr}. Then one would have to rethink the
 the implication of the 
current measurement of $n-1$ and of the current bound on $r=16\epsilon$,
for non-gaussianity in slow-roll inflation. Of course, such a rethink
is hardly going to alter the conclusion that 
non-gaussianity in this model will be very hard, if not impossible
\cite{cooray} to detect.

Now we turn to the  loop contributions.
 In the  presence of  cubic and higher terms, one finds
 integrations over a single momentum.
In the graphical representation, these correspond to  loops  which start
and finish at the same vertex.
It has been shown \cite{bksw} that, instead of including such loops,
one can replace the  factors $N'$, $N''$ and so on by
\be
N'\equiv N'(\sigbar)  \to  \vev{N'(\bfx)} \equiv \tilde N'
, \ee
and so on.\footnote
{The authors of \cite{bksw} use $\tilde N'$ to denote $N'(\bfx)$,
so that our $\tilde N'$ is equal to  their $\vev{\tilde N'}$.}
When these `renormalized vertices' are used, one need only draw `renormalized
graphs', which  omit all lines corresponding to an integration over a single
momentum. 
We are working at cubic order, which means that only $N'$
gets renormalized: 
\bea
N'(\bfx) &=&  N' + N'' \dsig(\bfx) + \frac12 N''' \dsig^2(\bfx) \\
\tilde N' &=&  N' + \frac12 N''' \vev { \dsig^2 }
, \eea
with
\be
\vev { \dsig^2 } = \int^{\ksmooth}_{\lmone} \calps(k) \frac{dk}k
\simeq \calps \ln(\ksmooth L)
. \label{dssq} \ee

Let us  verify that  the  loop contributions to the spectrum are 
still suppressed  at cubic order.
The  renormalized  tree-level contribution is
\be
\tilpps\su{tree} = \tilde N'^2 P_\sigma(k) \label{ptree}
. \ee
Using \eq{dssq} and the bound \eq{tree3} we find 
\be
\frac{\tilde \calpz\su{tree} - \calpz\su{tree} }
{\calpz\su{tree}} \sim \frac{N'''}{N'^3} \calpz \vev{\dsigs}
\lsim 10^{-5}
. \ee
With the cubic truncation we have 1- and 2-loop contributions, which 
have no renormalization. The 1-loop contribution to $\calpz$ 
 given by \eq{psigs}. The 2-loop contribution is \cite{bksw}
\be 
\ppz\su{2-loop} 
= \frac16 \frac{N'''^2}{(2\pi)^6} \int_{\lmone} d^3q_1 d^3q_2
P(q_1) P(|\bfq_2-\bfq_1|) P(|\bfk- \bfq_2|) \label{p2loop}
. \ee
Taking the integral to be dominated by the three singularities,
we estimate
\bea
\ppz\su{2-loop} &\sim& \frac{N'''^2}{N'^2} \calpz
\int_{\lmone} d^3p \pps(p) \pps(|\bfp-\bfk|) \\
&\sim & \frac{N'''^2}{N'^2N''^2} \calpz \ppz\su{1-loop} \\
&\sim& \( \frac{N''' \calpz}{N'^3} \)^2 \ppz\su{tree} \lsim
10^{-10} \ppz\su{tree} .
\eea

\section{Running}

We have advocated the use of a minimal box, but we did see that it might
be useful to consider also a super-large box in order to get a handle
on $\sigbar$ within a minimal box. 
Suppose that for some reason we decide to perform the whole calculation of 
 the correlators in   some super-large  box with size $L$. We may compare
the outcome of such a calculation with one done in some smaller box 
with size $M\ll L$, placed within the super-large box.
(I am thinking  of the size $M$ as being minimal but that is not essential.)
 An interesting
situation then arises, which was explored for the quadratic case in 
\cite{bl,mycurv}.\footnote
     {In \cite{mycurv} the labels $L$ and $M$ are interchanged so 
 that $L<M$. In \cite{bl} the labels are as here, but 
in Eq.~(24) of \cite{bl} it should be $\log(kM)$ instead of $\log(kL)$.}
I now  repeat  that analysis in a different way, arriving at
a differential equation instead of a finite-difference one, and extend
it to the cubic truncation. Then I ask how the calculation may be of practical
importance.

\subsection{General situation}

I think of the super-large box as having a fixed
size $L$. For a calculation within the a  smaller box of given size
$M$  and a given location,   $\sigbar$ of the previous equations 
becomes $\sigbarm$, 
and \eqs{withlin}{withlindefs} become
\be \zeta(\bfx) =  
 N_M' \dsigm(\bfx)  + \frac12 N''_M \dsigm^2(\bfx) + 
\cdots , \label{withlinell} \ee
where  
\be \dsigm(\bfx) = \sigma(\bfx) - \sigbarm
, \ee
and 
\be N'_M= \left. \frac{d N}{d \sigma} \right|_{\sigbarm}, \qquad
N''_M= \left. \frac{d^2 N}{d \sigma^2} \right|_{\sigbarm}
, \ee
and so on. In terms of the original quantities we have
\bea \dsigm &=& \( \sigbar - \sigbarm \)+ \dsig(\bfx)  \\
N'_M &=& N' + N''\( \sigbarm - \sigbar \) +\frac12 N''' 
\( \sigbarm - \sigbar \)^2 + \cdots
. \label{nell} \eea

For a calculation within the smaller box, 
correlators are   defined (in position space) as 
averages over the smaller box. The spectrum  of $\dsig$ is not
affected, because $\dsigm$ differs from $\dsig$ only
by a constant.
If $\dsig$ is gaussian, we can therefore forget about the change
in box size as far as its stochastic properties are concerned.
The same is not true of the  correlators of $\zeta$ though;
they will be different because $\zeta(\bfx)$ is a different
function, and because the average is taken over a smaller region.
Let us denote the spectrum defined within the smaller box by
$\ppzm$ and similarly for the bispectrum and higher correlators.

If we fixed the  size and location of the smaller box  within the original
 box, that would be the end of the story. But if the smaller box 
surrounds  the observable Universe, it may be reasonable to  suppose  that
we  occupy a typical position within the original box.
In that case, instead of considering $\ppzm$ and so on, one might
consider $\vev{\ppzm}$ and so on, the quantities obtained by 
averaging over the location of the smaller box
 while keeping its
size $M$ fixed. One might hope that this average will give  a reasonable
estimate of the correlators, evaluated within a smaller box of size $M$ that is 
fixed at our unknown location.

Since the correlators calculated within a given box 
can be defined as spatial averages within that box, $\vev{\ppzm}$ and so
on must be equal to the quantities $\ppz$ and so on, evaluated directly
within the super-large box. 
However, if  $\ppzm$ and so on are 
 calculated from \eq{withlinell} and the spatial
average within the super-large is then taken, the
 separation into
 tree-level and loop contributions is different. The loop contributions
will increase  with $M$, and the tree-level contributions will fall  to 
compensate. As we noticed earlier, the tree-level contribution will
usually dominate if the size $M$ is minimal, but that need not remain
the case as $M$ is increased to eventually become equal to the
super-large box size $L$. In the following sections we see
how the compensation occurs, first for the quadratic truncation and then
for the cubic truncation.

The cosmological  situation that we have described is
analogous to one that occurs in quantum field theory.
There,  one also calculates correlators (usually time-order, corresponding
to scattering amplitudes) which are the sum of a tree-level and loop
contribution. To do the calculation one has to specify a renormalization
scale $Q$.
The correlators are independent of $Q$ but the separation into
tree-level and loop contributions is
not. By choosing $Q$ to be of the same order as the relevant energy scale 
(set say by the momenta in a scattering process) the tree-level contribution
will normally dominate if it is present, otherwise the one-loop contribution
will normally dominate and so on. The cosmological situation that we consider
is similar, with $Q$ replaced by $M$. The maximal value $L$ of the super-large
box, determined by the amount of slow-roll 
inflation long before the observable Universe leaves
the horizon, provides the infrared cutoff of the theory. Its field theory
analogue is the ultra-violet cutoff of the effective field theory,
that dictates the maximum choice of the renormalization scale $Q$.

\subsection{Quadratic truncation}

Because \eq{withlinell} is only quadratic, $N''$ is just a number and
the expectation value with respect to the super-large box 
is required only for the 
tree-level terms. We have
\bea
\vev{\ppz\su{tree}} &=& \vev{N_M'^2} \pps(k)  \\
\ppz\su{loop}&=& 
\frac{N''^2}{(2\pi)^3}    \int_{\mmone} d^3p \pps(p)\pps(|\bfp-\bfk|) \\
\vev{ B_{\zeta}\su{tree} } &=& 2 \vev{ N_M'^2} N''
  \pps(k_1)  \pps(k_2) + { \rm cyclic} \\
B_{\zeta}\su{loop}&=&  \frac{N''^3}{(2\pi)^3}   \int_{\mmone}
d^3p  \pps(p)\pps(p_1) \pps(p_2) \\
\vev{ T_{\zeta}\su{tree} } &=&  \vev{N_M'^2}  N''^2
  \calp_\zeta(k_1) \calp_\zeta(k_2) \calp_\zeta(k_{14}) 
+  23 {\rm perms.}  \\
T_{\zeta}\su{loop}&=&  \frac18  \frac{N''^4}{(2\pi)^3}  \int_{\mmone}  d^3p 
\pps(p)\pps(p_1) \pps(p_2)\pps(p_{24}) \nonumber \\
&+&  23 {\rm perms} 
. \eea

Using \eq{nell} we find
\be
\vev{N'^2_M} = N'^2 + N''^2 \vev{(\sigbarm-\sigbar)^2}
. \ee
Considered as a function of the position of the  box with size $M$,
$\sigbarm-\sigbar$ is simply $\dsig(\bfx)$
smoothed with a top-hat window function. Its mean-square is therefore
\be
\vev{(\sigbarm-\sigbar)^2} = \int^{\mmone}_{\lmone}
 \calps(k) \frac{dk}k 
. \label{sigbarm} \ee
This expression generates $M$-dependence 
\be
\frac d {d\ln M}  \vev{(\sigbarm-\sigbar)^2} = - \calps(\mmone)
\label{diffform} . \ee 
The $M$-dependence of the tree-level contribution to the  spectrum is therefore
\be
\frac{ d\ppzm\su{tree}}{d\ln M} = -N''^2 \pps(\mmone)\pps(k) 
, \ee
and similar expressions hold for the  other tree-level contributions.

Now we come to the loop contributions. 
Because physical momenta have  $k\gg\lmone$, we can set $P(p)
= P(\lmone)$ near a  singularity of the integral at $p=0$.
Differentiating the integral with respect to $L$, we 
see that  the required cancellation occurs:
\be
\frac{d  }{d\ln M} \ppz\su{tree}
= -\frac{d  }{d\ln M}  \ppz\su{loop}
. \ee
Similarly, taking into
account all of the singularities, one can see that the same is true
for the bispectrum  and spectrum.

Returning to the analogy with 
quantum field theory, \eq{diffform} is like the `running' of a coupling
constant (or other parameter) with the renormalization scale $Q$. This running
makes the correlators of the field theory independent of $Q$.

\subsection{Cubic truncation}

I will just consider the spectrum. Evaluating it in  the smaller box and 
then taking the expectation value with respect to the original box,
\eqst{ptree}{p2loop} become
\bea
\tilpps\su{tree} &=& \vev{\tilde N'^2_M} P_\sigma(k)  \\
\tilppz\su{1-loop} &=& 
\frac12\frac{\vev{N''^2_M} }{(2\pi)^3}    \int_{\mmone} d^3p
\pps(p)\pps(|\bfp-\bfk|) \label{p11oop} \\
\ppz\su{2-loop} 
&=& \frac16 \frac{N'''^2}{(2\pi)^6} \int_{\mmone} d^3q_1 d^3q_2 \nonumber \\
&\times& P(q_1) P(|\bfq_2-\bfq_1|) P(|\bfk- \bfq_2|)
. \eea

The renormalized vertex in the smaller box is
\bea
\tilde N'_M &=&  N'_M + \frac12 N''' \vev { \dsig^2 }_M \\
\vev{ \dsig^2 }_M &=& \int^{\ksmooth}_{\mmone} \pps(k) \frac{dk} k
. \eea
Using \eq{nell} this gives
\be
\vev{\tilde N'^2_M} = N'^2 + \( N''^2 + N' N''' \) \vev{(\sigbarm-\sigbar)^2}
+\frac14 N'''^2 \vev{\dsig^2}^2
 \ee
Only the middle term is $M$-dependent, giving the running
\be
\frac d{d\ln M} \vev {\tilde N'^2_M} = N''^2 \pps(\mmone)
. \ee
The running of the   renormalized tree-level contribution  
$\tilppzm\su{tree}$ is therefore simply
\be
\frac d{d\ln M} \tilppzm\su{tree} = -N'' \pps(\mmone)
\pps(k)
, \ee
the same as in the quadratic case. 

The  running of the prefactor of the renormalized one-loop contribution 
$ \tilppzm\su{1-loop}$ is
given by  \eq{nell} as
\be
\frac d{d \ln M} \vev{N''^2_M} = - N'''^2 \calp(\mmone)
. \label{lastrun} \ee
Taking into account the running of the integral calculated in the
quadratic case this gives
\bea
\frac d{dM} \tilppz\su{1-loop}(k)
&=&-\frac12 \frac1{\tpq} N'''^2 \pps(\mmone) \tilpps\su{1-loop}(k) \nonumber \\
&+&  N'' \pps(\mmone)\pps(k)
. \eea
Finally, the running of integral in $\ppz\su{2-loop}$ gives
\be
\frac d{dM} \ppz\su{2-loop}(k) = \frac36 \frac1{\tpq} N''^2 \pps(\mmone)
\tilpps\su{1-loop}(k)
. \ee
We see that the total running of  $\ppz$ vanishes as required.

\subsection{Application}

Does the running have a useful  application? At first sight the answer would
seem to be `yes', because by going down to a minimal box the loop 
contributions become negligible. Unfortunately, the gain is illusory
because we are not actually calculating correlators within any particular
minimal box. Instead we are calculation the expectation values of the 
correlators within a minimal box (taken within the super-large box).
But these are just  the
actual correlators calculated within the super-large box. 
As a result, the  calculation has all of the uncertainties, and possibly
fatal problems,  that come with the use of a super-large box.

 The problem is that the correlators calculated within a super-large
box may be quite  different from the ones
observed in our Universe. If we throw down a minimal box within the 
super-large one, the correlators calculated within the minimal 
box will depend on its location. There is no reason to think
that a particular correlator, calculated with the minimal box at our location,
will be very close to the result obtained by averaging the position
of the minimal box. There is even less reason to think  that such  will
be the case simultaneously for all correlators. 

To quantify this concern one would like an estimate of the likely difference
between the averaged correlator and the  one observed. Extending the 
terminology coined for the cmb multipoles, one may call that cosmic variance.
It will be defined by the correlators evaluated within the super-large box.

As an example we may consider  the  simplest curvaton model,
where the curvaton has a quadratic potential and $\dot H/H^2$ is negligible
(see \cite{mycurv} for this case and further references).\footnote
{Taken literally this case is not realistic because it gives spectral
index bigger than 1. Keeping $\dot H/H^2$ negligible, the 
required value $n=0.95$ could be generated by giving the quadratic potential
a bump in the middle.  Alternatively one could invoke
significant $\dot H/H^2$ though  the probability distribution
of $\sigbar_M$ would not then be given by \eq{fprob}.}
 In that case, $\sigbar$ vanishes in a super-large
box and $\sigbar_M$ is generated entirely by the perturbation $\dsig_M$,
having the gaussian probability distribution given by \eq{fprob}. 
 According to taste, one may fold in this {\em a priori} expectation with 
environmental considerations.

\section{Conclusion}

I have explained how the use of a minimal box leads to fairly clean 
predictions. In particular I have verified that it makes some specific  loop 
contributions small, by virtue of
observational constraints on non-gaussianity.

I have also pointed to some of the uncertainties and possibly fatal problems,
that may come with the use of a super-large box.  
Some of the issues raised
here are quite deep and more work needs to be done.
Provisionally though, it would seem that
the only use of a super-large box is in its possible provision of
a  probability distribution, for the average of a curvaton-type field 
within a minimal box. 

{\it Acknowledgments.}~ 
I thank Dmitri Podolsky for useful comments on an earlier version of the 
paper.
The research is supported by  PPARC grant  PP/D000394/1   and by EU grants
MRTN-CT-2004-503369 and MRTN-CT-2006-035863.

\end{document}